\documentclass[aps,prb,twocolumn,superscriptaddress,showpacs,amsmath,amssymb,footinbib,longbibliography]{revtex4-2}
\usepackage[english]{babel}
\usepackage{graphicx}
\usepackage{dcolumn}
\usepackage{bm}
\usepackage{subfigure}
\usepackage[utf8x]{inputenc}
\usepackage{color}
\usepackage{textcomp}
\usepackage[colorlinks,bookmarks=false,citecolor=blue,linkcolor=red,urlcolor=blue]{hyperref}
\newcommand{\op}[1]{%
    \fontdimen12\textfont3=2pt\fontdimen12\scriptfont3=1.4pt%
    \!\null\mathop{\vphantom{#1}\smash{#1}}\limits_{\sim}\null\!}
\newcommand{\xref}[1]{\protect\ref{#1}}
\newcommand{\figref}[1]{Fig.~\protect\ref{#1}}

\newcommand{\fmref}[1]{(\protect\ref{#1})}
\def\bra#1{\langle \, {#1} \, | \,}
\def\ket#1{\, | \, {#1} \, \rangle}
\newcommand{\braket}[2]{\langle \, {#1} \, | \, {#2} \, \rangle}
\newcommand{\Tr}{\mbox{Tr}}
\renewcommand{\eqref}[1]{Eq.~(\protect\ref{#1})}
\newcommand{\vecops}[1]{\op{\vec{s}}_{#1}}
\def\half{{\frac{1}{2}}}

\begin{document}
\title{Finite-temperature Lanczos for anisotropic spin systems using triple-hybrid high-performance computing}
\author{J\"urgen Schnack}
\email{jschnack@uni-bielefeld.de}
\affiliation{Fakult\"at f\"ur Physik, Universit\"at Bielefeld, Postfach 100131, D-33501 Bielefeld, Germany}

\begin{abstract}
Lanczos  concepts  are  often  realized  on  supercomputers. In view of modern architectures of
high-performance computing triple-hybrid schemes employing MPI, CPU-openMP as well as GPU-openMP
should be utilized. The present article sketches how such a scheme could be utilized in order to 
meet the demands of the finite-temperature Lanczos method for anisotropic spin systems.
\end{abstract}

\maketitle

\section{Introduction}
\label{sec-1}

Thermodynamic equilibrium expectation values can be accurately approximated 
using trace estimators together with Krylov space representations.
These methods are heavily employed for spin models as well as Hubbard models but
are not restricted to these. 
The trace is estimated by a simple evaluation of an expectation 
value with respect to a random vector
\cite{Ski:88,Hut:CSSC89,JaP:PRB94,WWA:RMP06},
i.e.,
\begin{eqnarray}
\label{E-1-1}
\text{tr}
\left(\op{A}\right)
&\approx&
\bra{r}\op{A}\ket{r}
\;
\frac{\text{dim}({\mathcal H})}{\braket{r}{r}}
\ .
\end{eqnarray}
Here, $\op{A}$ is the operator of interest, and $\ket{r}$ is a 
vector randomly drawn from a (potentially high-dimensional)
Hilbert space. 
The complex components $r_{\nu}$ of the vector $\ket{r}$ 
with respect to a chosen orthonormal basis $\{\ket{\nu}\}$,
\begin{eqnarray} \label{eq:te}
\label{E-1-2}
\ket{r}
&=&
\sum_{\nu}\;
r_{\nu} \ket{\nu}
\ ,
\end{eqnarray}
are supposed to follow a Gaussian distribution with zero
mean, but the method 
works robustly with other choices of random numbers \cite{SRS:PRR20}.

In equilibrium statistical quantum physics such traces concern 
operators as $\exp\{-\beta\op{H}\}$ and
$\op{O}\exp\{-\beta\op{H}\}$ yielding the partition function and thermal 
expectation values of observables $\op{O}$, respectively. 
For correlated electron systems magnetic observables are often evaluated this way, 
see e.g.\ \cite{DRdV:ZP89,JaP:PRB94,JaP:AP00,ADE:PRB03,ZST:PRB06,HaS:EPJB14,SHP:PRB16,ScT:PR17,OAD:PRE18,SSR:PRB18,RKK:PRB19,JWW:JPSJ21} 
for a small selection of references, 
but the method is also used elsewhere, see e.g.\ \cite{MHL:CPL01},
{and with modifications under the name thermal pure quantum states, see e.g.\
\cite{SuS:PRL12,SuS:PRL13}.}

In this paper, we focus on the finite-temperature Lanczos method (FTLM)
\cite{JaP:PRB94,JaP:AP00,PRE:COR17}, a variant of the above idea, 
and apply it to anisotropic spin systems. The Hamiltonian of such systems
might contain anisotropic exchange, dipolar interactions \cite{WRK:PRL18,MNK:CP24} or
single-ion anisotropies \cite{SWB:IC23} together with
isotropic terms such as Heisenberg exchange interactions.
It was noted earlier that the typically very good convergence
of FTLM for Heisenberg systems worsens drastically 
under anisotropy in particular
at low temperatures \cite{ADE:PRB03,Mun:WJCMP14,HaS:EPJB14,MoT:PRR20}.
We follow the numerically costly workarounds proposed in 
\cite{HaS:EPJB14,ADE:PRB03} and present 
an implementation for high-performance computing. With two examples 
we will demonstrate the applicability of FTLM also for very anisotropic 
spin systems.

The paper is organized as follows. In Section \ref{sec-2} we introduce 
model and methods. Results are discussed in Sec.~\xref{sec-3}.
The article closes with a discussion in Section~\ref{sec-4}.

\section{Model and methods}
\label{sec-2}

\subsection{Spin model}
\label{sec-2-1}

The Hamiltonian that is employed in the following consists of 
Heisenberg exchange interactions, single-ion anisotropies, 
the Zeeman term as well as {intramolecular} dipolar interactions
\begin{align}\label{eq:mce-hamop}
	\op{H}=&2 \sum_{i<j} J_{ij} \vecops{i}\cdot\vecops{j} 
	+ \sum_i d_i \left(\vecops{i} \cdot \vec{e}_i\right)^2
	+ \mu_B  \vec{B}\cdot \sum_i g_i \vecops{i}
	\\
	&+ \frac{\mu_0\mu_B^2}{4\pi}\sum_{i<j}\frac{g^2}{r_{ij}^3}\left(\op{\vec{s}}_i\cdot \op{\vec{s}}_j-3\cdot\op{\vec{s}}_i\cdot \vec{e}_{ij}\otimes\vec{e}_{ij}\cdot \op{\vec{s}}_j\right)
\nonumber
    \ ,
\end{align}
{where the first sum runs over ordered pairs of indices $i<j$.}
Here, $\op{\vec{s}}_j$ denotes the spin vector operator at site $j$, and 
a tilde is used to mark operators in general.
Single-ion anisotropy is incorporated in its simplest form with
anisotropy axes along local directions $\vec{e}_i$.

If the dimension of the related Hilbert space is sufficiently small
the respective matrix representation can be diagonalized numerically, 
if possible using symmetries, see e.g.\ \cite{Wal:PRB00,ScS:IRPC10,GhK:Mag25,WSU:SciPostPhysCodeb.70,WSU:SciPostPhysCodeb.70-r0.4,}. 
However,
since the dimension, $d=\prod_i (2s_i+1)$, grows exponentially with 
the number of spins, exact diagonalization is rather limited. This is
even more true for anisotropic spin systems since they 
typically lack symmetries. Then, approximate methods such
as FTLM are needed.

\subsection{Finite-temperature Lanczos method}
\label{sec-2-2}

The finite-temperature Lanczos method (FTLM)
\cite{JaP:PRB94,JaP:AP00,PRE:COR17} approximates an equilibrium 
observable as 
\begin{eqnarray}
\label{E-2-C}
O^{\text{FTLM}}(T,\vec{B})
&\approx&
\frac{\sum_{r=1}^R\;\bra{r}\op{O}e^{-\beta \op{H}}\ket{r}}
     {\sum_{r=1}^R\;\bra{r}e^{-\beta \op{H}}\ket{r}}
\ .
\end{eqnarray}
Here, the set of $R$ random vectors $\ket{r}$ is the same 
in numerator and denominator. If the Hamiltonian $\op{H}$ 
possesses symmetries, these can be 
employed by decomposing the full Hilbert space into
mutually orthogonal subspaces according to the irreducible
representations of the employed symmetry
\cite{ScW:EPJB10,HaS:EPJB14}
yielding for the partition function
\begin{eqnarray}
\label{E-2-D}
Z^{\text{FTLM}}(T,\vec{B})
&\approx&
\sum_{\gamma=1}^\Gamma\;
\frac{\text{dim}({\mathcal H}(\gamma))}{R}
\nonumber
\\
&&
\times
\sum_{r=1}^R\;
\sum_{n=1}^{N_L}\;
e^{-\beta \epsilon_n^{(r)}} |\braket{n(r)}{r}|^2
\ .
\end{eqnarray}
${\mathcal H}(\gamma)$ is the subspace belonging to 
the irreducible representation $\gamma$, $N_L$ is the dimension
of the generated Krylov space, and $\ket{n(r)}$ is the n-th
eigenvector of $\op{H}$ in this Krylov space with seed $\ket{r}$
and energy eigenvalue $\epsilon_n^{(r)}$. $N_L$ is typically chosen such,
that the ground state energy in the respective subspace is converged
to numerical accuracy, which often already happens for $N_L$ of
the order of 100 \cite{SRS:PRR20}.
$\beta=1/(k_B T)$ denotes the inverse temperature. 

The arguably most often employed symmetry is related to the conservation 
of the $z$-component of the total spin $\op{S}^z$ that is 
for instance present in Heisenberg
and XXZ models. Investigations of anisotropic systems where the $\op{S}^z$-symmetry
does not hold revealed that this symmetry is responsible for the 
astonishing accuracy and fast convergence of magnetic observables in FTLM. 
Both, accuracy as well as
fast convergence worsen massively if anisotropy breaking
$\op{S}^z$-symmetry is present -- the stronger the worse.
It was argued that both problems can be cured by making use of the always present
time-reversal symmetry by utilizing pairs of mutually time-reversed random vectors
for the FTLM approximation \cite{HaS:EPJB14}. We will follow this suggestion in the 
present paper.

A second concern is related to the asymmetry in \fmref{E-2-C}, 
which was first discussed in \cite{ADE:PRB03}. 
At the level of the trace we observe that
\begin{eqnarray}
\label{E-2-E}
\Tr\left[\op{O}e^{-\beta \op{H}}\right]
&=&
\Tr\left[e^{-\half \beta \op{H}}\op{O}e^{-\half \beta \op{H}}\right]
\ ,
\end{eqnarray}
but this is in general not true for the expectation value
\begin{eqnarray}
\label{E-2-F}
\bra{r}\op{O}e^{-\beta \op{H}}\ket{r}
&\neq&
\bra{r}e^{-\half\beta \op{H}}\op{O}e^{-\half\beta \op{H}}\ket{r}
\ .
\end{eqnarray}
The authors of \cite{ADE:PRB03} demonstrated a massively lowered 
accuracy for small temperatures if the asymmetric form 
\fmref{E-2-C} is employed, 
but depending on the anisotropy this problem 
is not restricted to low temperatures.
One also notices, that if $\op{O}$ commutes with $\op{H}$ the expressions
in \fmref{E-2-F} are the same which is at the heart of the good convergence 
for $\op{S}^z$-symmetric problems.

In the remainder of this paper we will thus employ the symmetric version
\begin{eqnarray}
\label{E-2-G}
O^{\text{sFTLM}}(T,B)
&\approx&
\frac{\sum_{r=1}^R\;\bra{r}e^{-\half\beta \op{H}}\op{O}e^{-\half\beta \op{H}}\ket{r}}
     {\sum_{r=1}^R\;\bra{r}e^{-\half\beta \op{H}} e^{-\half\beta \op{H}}\ket{r}}
\end{eqnarray}
and term it symmetric FTLM (sFTLM = low-temperature FTLM in \cite{ADE:PRB03}). 
Although looking rather innocent, 
this improvement comes at the cost of another factor $N_L$ 
in computing time because the expectation value
\begin{eqnarray}
\label{E-2-H}
&&\bra{r}e^{-\half\beta \op{H}}\op{O}e^{-\half\beta \op{H}}\ket{r}
\\
&\approx&
\sum_{m=1}^{N_L}\;
\sum_{n=1}^{N_L}\;
e^{-\half\beta (\epsilon_m^{(r)}+\epsilon_n^{(r)})} 
\nonumber
\\
&&\times
\braket{r}{m(r)}
\bra{m(r)}\op{O}\ket{n(r)}
\braket{n(r)}{r}
\nonumber
\ ,
\end{eqnarray}
now contains a double sum, compare to \fmref{E-2-D}. 
Since $N_L\sim 100$ this dramatically increases the computing time
\cite{RKK:PRB19}. In addition, all eigenvectors $\ket{n(r)}$ 
need to be stored.

\subsection{Triple-hybrid high-performance computing}
\label{sec-2-3}

Lanczos concepts \cite{Lan:JRNBS50} are often realized on supercomputers, see
\cite{SSR:PRB18,WiL:PRE18,LSM:PRB19,AAB:N19,NTS:JPSJ19,DJW:CPC19,SSH:PRL20,NTT:JPSJ22,WWF:CPC22,NaS:JPSCP:23}
for some of the largest more recent calculations.
The paradigm that high-performance computations should be programmed 
using a MPI/openMP hybrid scheme as employed for instance in the
publicly available program
\emph{spinpack} \cite{spin:258,RiS:EPJB10} moves towards 
triple-hybrid
MPI/CPU-openMP/GPU-openMP with recent updates of supercomputers
such as the Phase-2 at the Leibniz supercomputing center (LRZ)
of the Bavarian Academy of Sciences in Germany \cite{LRZ}.
Now, a program should be parallelized across MPI ranks, CPU threads, and
GPU engines where GPU is the graphics processing unit (\emph{vulgo} graphics card).
In the following, a working approach is presented for the sparse matrix 
vector multiplications that are at the heart of every Lanczos procedure.

The approach presented here distributes the Hamiltonian matrix and the
Lanczos vectors across the ranks of a MPI parallelization and takes care
of the sparsity at the level of the resulting small sub-matrices. 
In a practical realization a \emph{rank} would correspond to a GPU, i.e., 
for a system such as Phase-2@LRZ with nodes of 4 GPUs one would address
four times as many ranks as \emph{nodes}.
For technical clarity, a node denotes a pc-like part of a supercomputer,
a rank is one of the parallel MPI processes of which there are \emph{size} many. 

\begin{figure}[ht!]
\centering
\includegraphics*[width=0.75\columnwidth]{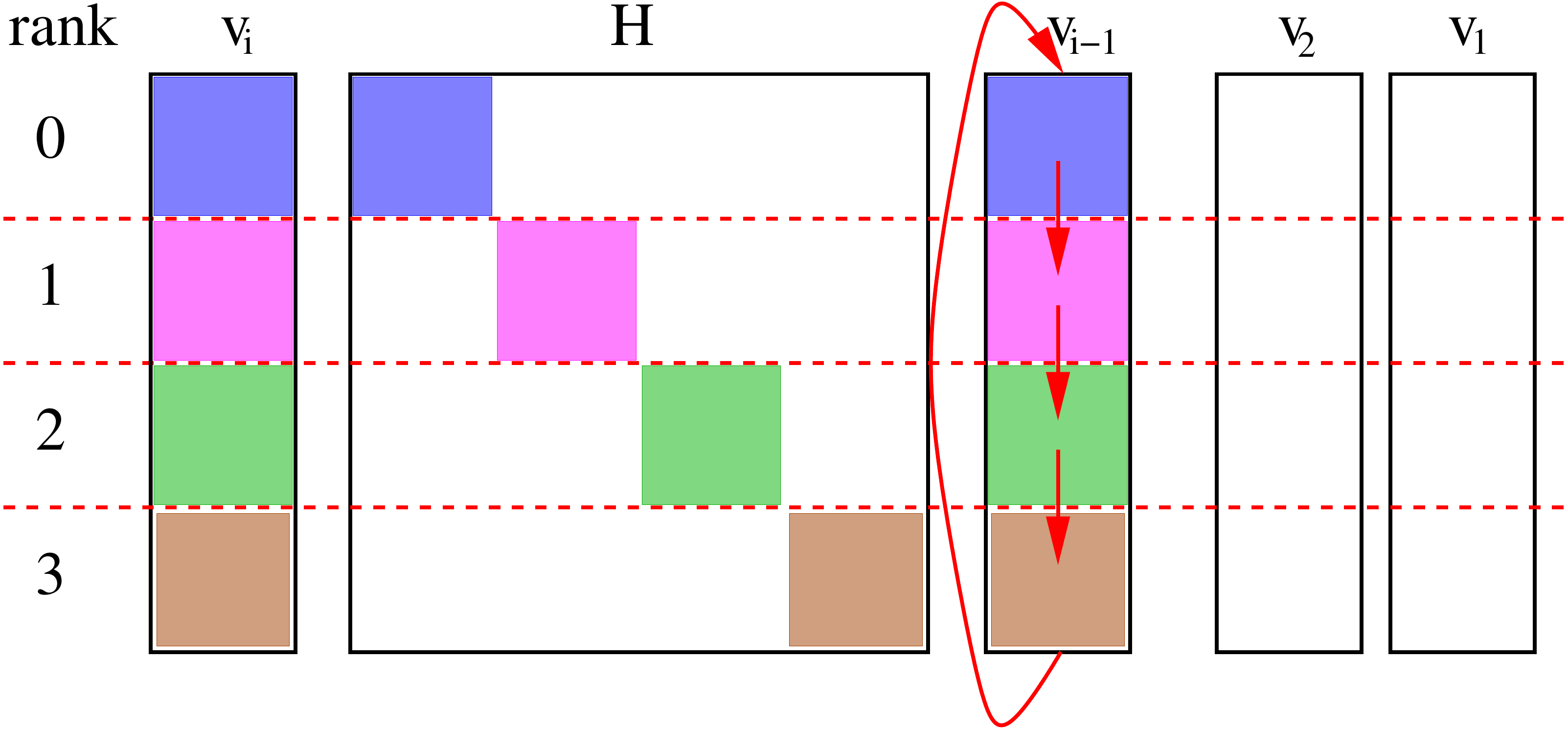}
\caption{Sketch of the MPI parallelization of a matrix-vector multiplication. 
Each rank hosts a section of all vectors and a stripe of the matrix. The first step fills 
the initially blank target vector $v_i$ with the contributions of the respective sections 
of $v_{i-1}$ on the same rank. The needed parts of the matrix are given by colored
blocks.}
\label{sFTLM-f-1}
\end{figure}

In a first step, the Lanczos vectors used for the buildup of the Krylov space
will be split into as many (almost) equal sections as there are ranks,
i.e., $ranks = 0, 1, \dots, size-1$.
How many  Lanczos vectors are stored depends on 
whether one wants to perform reorthogonalization 
against earlier vectors. The minimal number is two predecessor
vectors to construct the new Lanczos vector.
In a similar manner the Hamiltonian 
matrix is split into as many slices, see \figref{sFTLM-f-1}.
The idea is to rotate the sections of one of the vectors among the MPI ranks
and leave the matrix stripes stationary where they are created. In the scheme 
presented in \figref{sFTLM-f-1} the source vector $v_{i-1}$ will be rotated.
Thus the stripes of the Hamiltonian presented in the figure belong to $H$.
It is as well possible to rotate the target vector ($v_{i}$), then the stripes 
would belong to $H^T$.

Figure \xref{sFTLM-f-1} shows the first MPI-step of the matrix vector multiplication
that is in parallel executed on each rank. The initially blank vector $v_i$ 
is filled section-wise on each rank.

\onecolumngrid

\twocolumngrid

\begin{figure}[ht!]
\centering
\includegraphics*[width=0.75\columnwidth]{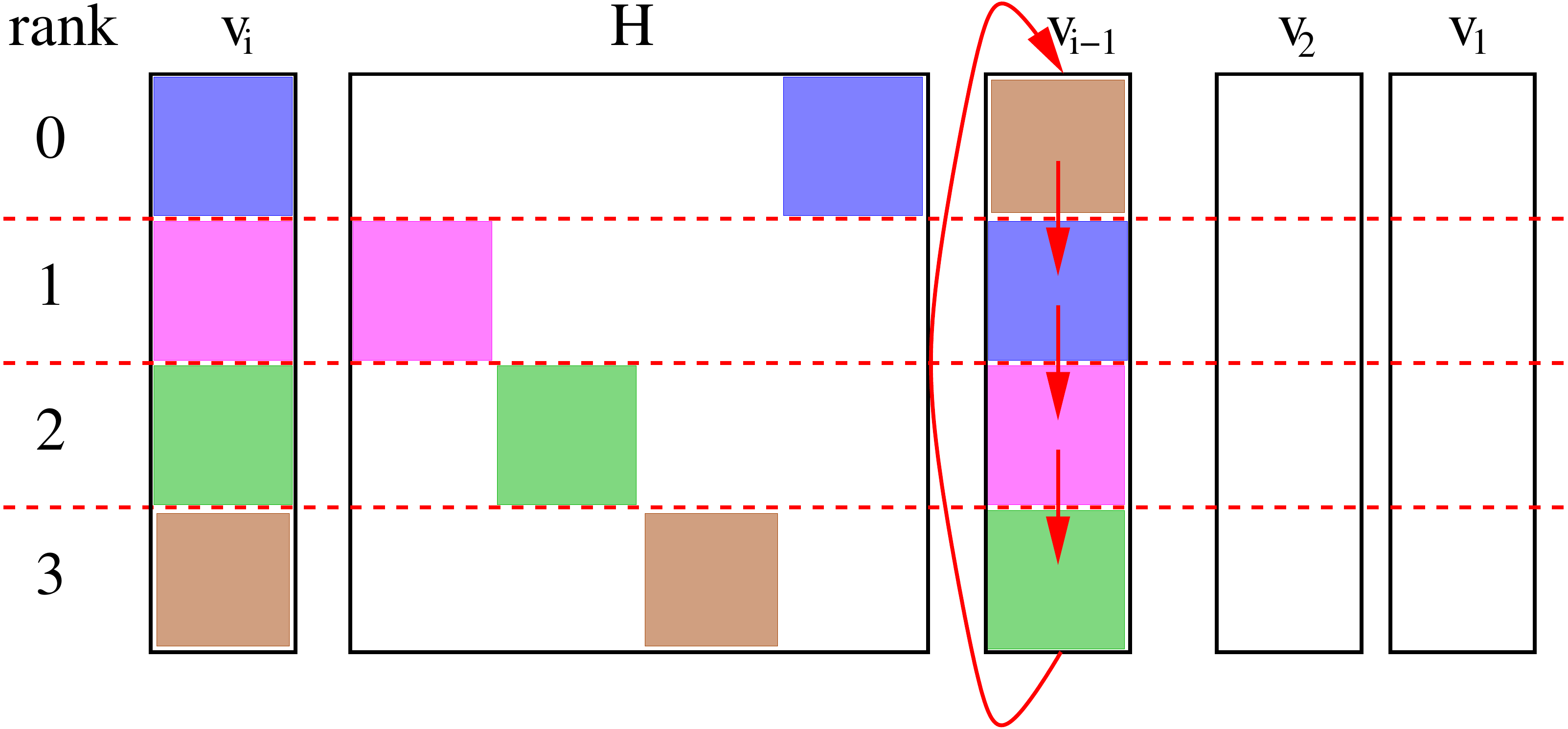}
\caption{Second MPI-step of the matrix-vector multiplication; the 
contributions of the respective sections of $v_{i-1}$ on the same rank are added by 
using the correct part of the matrix stripe given by the colored boxes.}
\label{sFTLM-f-2}
\end{figure}

For the next MPI-step of the matrix-vector multiplication the sections of the source vector 
$v_{i-1}$ are rotated using \emph{mpi\_sendrecv} 
which yields the configuration shown in \figref{sFTLM-f-2}. Now, the 
contributions of the respective sections of $v_{i-1}$ on the same rank are added by 
using the correct part of the matrix stripe given by the colored boxes. This process is
continued until the vector $v_{i-1}$ has completed a full cycle and is again 
of the structure shown in \figref{sFTLM-f-1}.

In a Lanczos procedure $v_{i-1}$ and $v_{i-2}$ have to be projected out of $v_i$,
and for complete reorthogonalization all previous $v_j$. The advantage of the organization 
in stripes is that this can be done in parallel on every rank without any communication with 
other ranks (since we use an orthonormal basis). Only for the normalization 
of $v_i$ we need an \emph{mpi\_allreduce} command to add up the contributions of every rank.

\onecolumngrid

\twocolumngrid

The multiplication of the submatrices with the respective sections of the vector is 
done on the GPU. For this purpose, the matrix stripes should be set up and 
then (permanently) stored on the GPU 
using \emph{omp target teams} pragmas
so that only the processed  vector sections need 
to be uploaded to the GPU and the resulting sections downloaded, respectively.
To this end, the needed space is allocated on the GPU, the sparse matrix 
(the full stripe of $H$ on each rank) is uploaded only once at the beginning,
and the vectors on each rank are only updated using 
\emph{omp target update to/from}
pragmas.

For an easy manipulation, every stripe of H is subdivided into \emph{size} blocks 
(colored boxed in Figs.~\xref{sFTLM-f-1} and \xref{sFTLM-f-2}) 
by transforming the two-dimensional stripe into a sequence of blocks,
i.e., by reorganizing the two-dimensional array into three dimensions.
Each of these matrix blocks is then stored as sparse matrix.
We employ the COO (coordinate format) storage mode, i.e., store each sparse matrix as
three one-dimensional arrays containing indices $m$ and $n$ of 
the non-zero matrix elements $H_{mn}$. 
Hermiticity is not employed by storing only matrix elements with
$m\leq n$ since this would slow down the openMP loops for the matrix-vector
multiplication on the GPU.

On the GPU the matrix vector multiplication is done by employing
openMP for the loop about the entries of the three vectors. 
This would lead to \emph{race conditions} due to the multiple appearance 
of the same index of the target vector in COO format, i.e., 
it would lead to wrong assignments of entries of $v_i$
which would worsen the stability problems inherent in Lanczos
procedures \cite{EPS:PRB25}.
The problem could be cured by using the compressed sparse row (CSR) format,
where a parallelization over row pointers would prevent the race condition,
however, we used an alternative approach of grouping
the indices into disjoint sets as shown in the example in 
Table~\xref{sFTLM-t-1}. 
This also simplifies the parallel creation of the matrix in COO format.

\begin{table}[ht!]
\caption{\label{sFTLM-t-1}%
Example on how the sparse matrices are stored using 
the auxiliary index iCore to structure the respective 
vectors into blocks of disjoint indices $m$. 
}
\begin{ruledtabular}
\begin{tabular}{rrrrrd}
\textrm{glob. index}&
\textrm{iCore}&
\textrm{loc. index}&
$m$ &
$n$ &
\multicolumn{1}{c}{$H_{mn}$}\\
\colrule
1 & 1 & 1 & 1 & 5 & 30.1\\
2 &   & 2 & 1 & 7 & -8.9\\
3 &   & 3 & 2 & 3 & 1.8\\
4 &   & 4 & 2 & 7 & -2.5\\
5 &   & 5 & 2 & 9 & 3.4\\
\colrule
6 & 2 & 1 & 3 & 2 & 1.8\\
7 &   & 2 & 3 & 5 & -15.6\\
\end{tabular}
\end{ruledtabular}
\end{table}

In Table~\xref{sFTLM-t-1}, the global index enumerates the non-zero matrix elements as well as 
the entries of the index vectors which in a simple openMP parallelization
about the global index would lead to race conditions for identical $m$. 
This is cured with the help of an auxiliary index $iCore$ that groups
indices into disjoint sets so that identical $m$ are treated one after 
the other since the GPU-openMP parallelization runs about the index $iCore$.

\section{Results}
\label{sec-3}

\subsection{HPC}
\label{sec-3-1}

The summary of the above is that the outlined programming scheme works. 
The program was compiled using the Intel\textsuperscript{\textregistered}
oneAPI\textsuperscript{\textregistered} suite, and it
runs on flexible numbers of ranks as tested on the Phase-2 supercomputer at LRZ 
employing nodes with two
Intel\textsuperscript{\textregistered} Xeon\textsuperscript{\textregistered} Platinum 8480+ 
CPUs (Sapphire Rapids)
and four
Intel\textsuperscript{\textregistered} Data Center GPUs Max 1550 (Ponte Vecchio).

A few characteristics can be deduced from the test runs performed at the LRZ 
and on a local AMD machine.
The total linear dimension of the Hilbert space is $(2s+1)^N$ for $N$ equal spins,
and the number of non-zero matrix elements per row is of order $o(N^2)$ since the dipolar
interaction connects all spins pairwise. However, the number of non-zero elements 
for each block of a stripe of the matrix
in Figs.~\xref{sFTLM-f-1} and \xref{sFTLM-f-2} can be very different which leads
to an imbalance.
For large projects with several nodes (up to 64 tested) MPI transfer times seem to
dominate the run time. The time to load vectors to and from the GPU is small in 
comparison \footnote{Thanks to help by Tobias Kl{\"o}ffel (Intel).}.
In the scheme employed so far additional idle times occur since each node 
has got many CPU cores which wait while the GPU is dealing with the matrix.
The same is true if the CPU cores work {to evaluate the matrix elements of the
magnetization}, and the GPU waits.
Our experience collected so far suggests that one should fill the nodes with as
much matrix and vector data as allowed by the available RAM. 
This reduces all communication times. 

Realistic future improvements could be achieved by processing 
several random vectors at the same time, for instance the two time-reversed vectors,
in order to attenuate random memory access on the GPU due to unstructured indices 
$n$ in the sparse storage format.
This would better the ratio of run time to load time of the GPU. 
In addition, one could alternate between vectors, i.e., 
process one group on the ranks while the other is MPI-transferred between ranks
and vice versa.
A more intricate improvement would be the parallel use of CPUs and GPUs.

\subsection{FTLM for anisotropic spin rings}
\label{sec-3-2}

The scientific goal of the outlined numerical procedure is to model 
anisotropic spin systems by means of the approximate finite-temperature Lanczos procedure 
in cases where they are too large for exact diagonalization of the 
Hamiltonian and also too large for an FTLM approximation on
a local workstation even with GPU \cite{TaK:A26}.
The example spin system discussed in the following possesses
a rather large single-ion anisotropy 
as e.g. in the case of dysprosium, manganese, or cobalt, see \cite{BDA:CAEJ:17} for
an example. In addition, anisotropic 
exchange might occur such as dipolar interactions. In order to estimate 
the accuracy and convergence of the above scheme we restrict our investigation
to a fictitious ring molecule with $N=6$ spins of spin quantum number $s=5/2$ that can
irrespective of the terms in the Hamiltonian be diagonalized numerically exactly. 
The Hamilton operator as given by \eqref{eq:mce-hamop} contains a nearest-neighbor 
Heisenberg exchange $J=0.02$~K, strong non-collinear single-ion anisotropy $d_i=-20$~K 
as well as all-to-all dipolar interactions. The single-ion anisotropy is dominant 
in the chosen example; the parameters are motivated by dysprosium ring
molecules as e.g.\ in \cite{BDA:CAEJ:17}, but the specific values do not 
really matter for the statements made in the following.

\begin{figure}[ht!]
\centering
\includegraphics*[width=0.90\columnwidth]{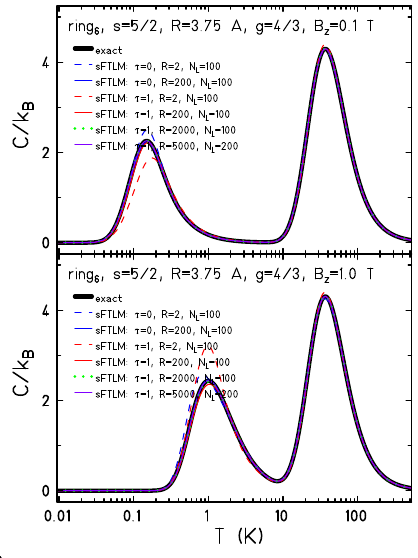}
\caption{Heat capacity of a fictitious spin ring of $N=6$ spins with $s=5/2$
with nearest-neighbor Heisenberg exchange, strong single-ion easy-axis anisotropy 
as well as all-to-all dipolar interaction for $B=0.1$~T (top) and
$B=1$~T (bottom). $\tau=1$ or $0$ denotes the use
or not-use of time reversed random vectors; 
$N_L$ provides the number of Lanczos steps
for a single random vector, and $R$ gives the number of employed random vectors
according to \fmref{E-2-G} and \fmref{E-2-H}.}
\label{sFTLM-f-3}
\end{figure}

Figure~\xref{sFTLM-f-3} shows as a first result the heat capacity for two
values of the magnetic field applied along the $z$-axis which is the symmetry
axis of the ring. The various curves origin from using time reversed random vectors
($\tau=1$) or not ($\tau=0$) together with $N_L$ Lanczos steps for each random vector
and averages of the partition function over $R$ random vectors
according to \fmref{E-2-G} and \fmref{E-2-H}. One notices that the heat capacity
is astonishingly robust and well approximated by already rather small 
numbers of random vectors and Lanczos steps of the order of 100. This even holds
for the low-temperature peak of the heat capacity. Only for single-figured numbers
of random vectors the accuracy is visibly lower, 
but only around the low-temperature 
peak. We conjecture that the reason is given by the fact, that the approximate 
energy eigenvalues converge much faster than the eigenstates as is known from
perturbation theory. Therefore, the heat capacity, which depends only on the 
energy eigenvalues, appears rather accurate and robust with smaller effort. 
This changes drastically for the magnetization.

\begin{figure}[ht!]
\centering
\includegraphics*[width=0.90\columnwidth]{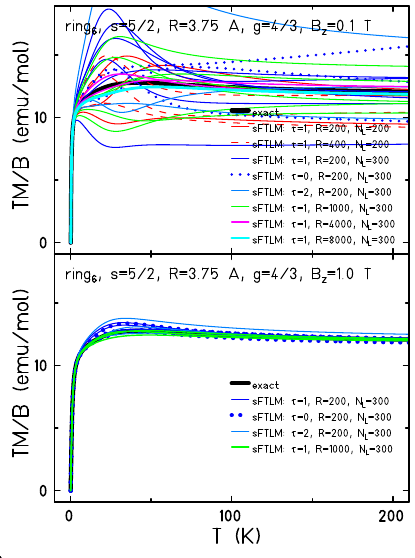}
\caption{Susceptibility, $M T/B$, of a fictitious spin ring of $N=6$ spins with $s=5/2$
with nearest-neighbor Heisenberg exchange, strong single-ion easy-axis anisotropy 
as well as all-to-all dipolar interaction for $B=0.1$~T (top) and
$B=1$~T (bottom). 
$\tau=1$ or $0$ denotes the use
or not-use of time reversed random vectors; 
$N_L$ provides the number of Lanczos steps
for a single random vector, and $R$ gives the number of employed random vectors
according to \fmref{E-2-G} and \fmref{E-2-H}. {Curves with the same color
show realizations of the same parameter set that just differ by the seed
of the random number generator. The exact black curve in the bottom panel 
is right below the green curves.}}
\label{sFTLM-f-4}
\end{figure}

The magnetization, displayed in \figref{sFTLM-f-4}, turns out to be much more delicate,
a behavior that can be traced back to the fact that now not only the approximate
energy values $\epsilon_m^{(r)}$ matter, but also the matrix elements of
the magnetization, compare \fmref{E-2-H}.
In addition, the typical presentation of susceptibility times temperature
magnifies high-temperature inaccuracies. Such inaccuracies are more
prominent for smaller values of the magnetic field. In such cases,
the Zeeman term might be of the same order as FTLM accuracy fluctuations 
in energy and weights.
On top of this, fluctuations of the magnetization are enlarged by
dividing by a small field $B$.

This all changes for the larger field of $B=1$~T 
(\figref{sFTLM-f-4}, bottom)
which converges much more
quickly for smaller numbers of both random vectors and Lanczos steps. In addition, 
the explicit use of time-reversal symmetry is no longer needed.

The various scenarios shown in \figref{sFTLM-f-4} can be summarized as follows.
Small numbers of random vectors up to at least $R=200$ lead
to wrong results, in particular for the case of 
small fields such as $B=0.1$~T (top panel).
The symmetric cases with $R=200$ and $\tau=1$ (blue curves) are somewhat better,
but still wrong. Fortunately, even without the correct reference untrustworthy curves 
can be identified since they, e.g., do not approach the paramagnetic limit. 
The finer details, for instance where the curves bend, converge rather
slowly in accordance with the observations in Ref.~\cite{ADE:PRB03}. 
It came as a surprise that even for $R=8,000$ (cyan curve) 
the approximation does not converge to
the exact result for the small field of $B=0.1$~T and lower temperatures
(top panel).

\subsection{Mn$_9$ -- a double pyramid}
\label{sec-3-3}

As an encouraging outlook we would like to present an example
where the authors stated in their recent article that a numerical treatment
of this spin system appears impossible to them, see \cite{BKV:ICF26}.

\begin{figure}[ht!]
\centering
\includegraphics*[width=0.90\columnwidth]{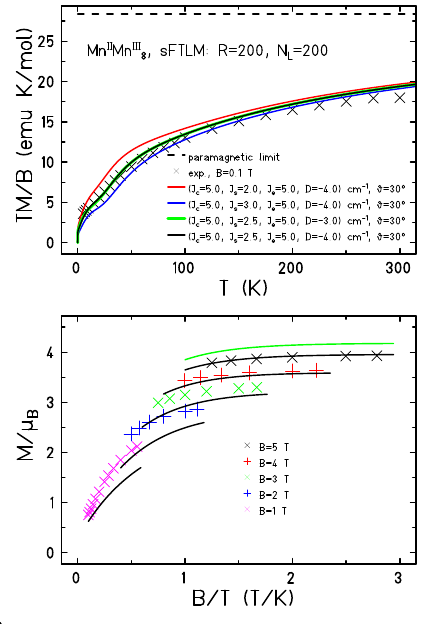}
\caption{Susceptibility, $M T/B$, of Mn$_9$. 
$N_L$ provides the number of Lanczos steps
for a single random vector, and $R$ gives the number of employed random vectors
according to \fmref{E-2-G} and \fmref{E-2-H}.
{Experimental} data taken from \cite{BKV:ICF26}.}
\label{sFTLM-f-5}
\end{figure}

We model Mn$_9 = $Mn$^\text{II}$Mn$^\text{III}_8$ as a centered
cube with three exchange interactions: $J_c$ between the central
Mn$^\text{II}$ of spin $s=5/2$ and the eight Mn$^\text{III}$
of spin $s=2$ at the vertices of the cube; $J_s$ between nearest 
neighbors on the top and bottom squares, and $J_e$ along the edges 
between top and bottom Mn$^\text{III}$. In addition, Mn$^\text{III}$
possesses an easy axis of strength $D$ that is tilted by $30^\circ$
with respect to the symmetry axis through top and bottom square 
and directed along $\phi=\pm 10^\circ$ of the $x$-axis. 
For details see \cite{BKV:ICF26}. We use powder averages about 
the vertices of a dodecahedron.

Since this is a paper on numerical methods we did not aim at a 
perfect fit, but rather want to demonstrate that with the 
proposed numerical scheme large anisotropic spin systems can
be modelled. Therefore, we picked a few reasonable parameter set guided
by our collected experience with Mn-based hourglas molecules
\cite{HGM:CS12,HHK:DT12,GHG:CCR15}. Our best results, green
and black curves in \figref{sFTLM-f-5}, show that Mn$_9$ 
with Hilbert space dimension of 2,343,750 can be modelled
with resonable effort. We used 4 nodes on SuperMuc Phase 2
with 16 GPUs in total. For each field value and 10 directions the
calculation took about 100 minutes.
The convergence of this system is better than for the ring
system discussed before since the single-ion anisotropy is much smaller
compared to exchange and field. Therefore, the low-energy density of
states is denser which enhances the convergence of FTLM.

If one would like to improve the agreement of theory and data,
then more refined and maybe less symmetric parameterizations,
compare\cite{BKV:ICF26}, would have to be investigated.

\section{Discussion}
\label{sec-4}

In view of the necessary large number of random vectors of the order of a few
thousand as well as $N_L\approx 200\dots 300$ Lanczos steps for each of them, 
the use
of massively parallel schemes is indispensable. A Hamiltonian matrix is applied
$R\cdot N_L$ times, i.e., a hundred thousand times -- this task is indeed very 
well suited for a GPU \cite{WWF:CPC22,TaK:A26}

There are a few possible future improvements that one could and should investigate.
One option might be an improvement of the storage mode of the Hamiltonian
matrix, compare Table~\xref{sFTLM-t-1}, and connected to this a more 
cache-efficient reading of the source vector on the GPU, e.g., 
by manipulating a group of vectors at the same time.

Another possible improvement would be a simultaneous use of GPUs and CPUs
which are both idle while the other works. 
However, an obstacle is always given by 
the fear to develop too specialized code that does not run elsewhere
or on potentiall different near-future architectures.

{A first test of GPU-aware MPI, i.e.\ MPI communication between GPUs
without involvement of CPU hints at a reduction of computing time by about
50~\%.}

{Since the sparse Hamiltonian matrix is stored in the GPUs the achievable size 
of quantum spin systems is still limited compared to schemes where matrix elements
are evaluated on the fly. An interesting molecule consisting of 8 nickel and 4 gadolinium 
ions could not be simulated on the available hardware \cite{LRZ}. We thus estimate
that systems with total Hilbert space dimension of the order of about
25,000,000 are already too big for the proposed scheme. For such and larger cases
a variant that evaluates matrix elements on the fly has to be developed.}

A final improvement concerns the rate of convergence of the Lanczos
procedure when approximating a densitity of states that exhibits 
pronounced bunching. Here a representation by Chebyshev polynomials 
as proposed in \cite{WWA:RMP06} might be the better approach,
however, without smoothing kernel as we demonstrated in 
\cite{SGS:ZNA21}.

\section*{Acknowledgment}

I would like to thank my group at Bielefeld University for many
fruitful discussions and for the patience with which they did stand 
my enthusiastic but long musings on matrices, MPI, openMP, and FORTRAN. 
I would like to thank Hans-Hermann Frese for a good introduction to MPI.

I would also like to thank Martin Ohlerich of the Leibniz supercomputing center (LRZ)
of the Bavarian Academy of Sciences for very many insightful discussions
on the various computational problems and strategies.

Thanks go also to J{\"o}rg Schulenburg (Magdeburg University), 
Hans De Raedt and Fengping Jin (FZ Jülich), 
Momme Allalen (LRZ) as well as 
Christopher Danken and Tobias Klöffel (Intel) for discussions on HPC of 
matrix vector multiplications.

Computing time at the LRZ trough projects pn39ga \& pn76ta is gratefully acknowledged.


%

\end{document}